# Impact of Nuclear Deformation of Parent and Daughter Nuclei on One Proton Radioactivity Lifetimes

**A. Jain**[1], **Pranali Parab**[2], **G. Saxena**[3,*], **and Mamta Aggarwal**[2]

[1]Department of Physics, School of Basic Sciences, Manipal University Jaipur, Jaipur-303007, India
[2]Department of Physics, University of Mumbai, Vidhyanagari, Mumbai, 400098, Maharashtra, India
[3]Department of Physics (H&S), Govt. Women Engineering College, Ajmer-305002, India
*gauravphy@gmail.com

**ABSTRACT**

The influence of nuclear deformation on proton-decay half-lives has been systematically studied in microscopic theoretical frameworks for a wide range of nuclei with Z<82. Correlation between 1p-decay half-lives and the deformed nuclear shapes of both the parent and daughter nuclei has been investigated. Since the deformations of proton emitters and their residual nuclei impact the potential barrier and disintegration energy which are crucial for the accurate determination of half-lives, we incorporate the nuclear deformations of both the emitters and residues in a phenomenological manner and propose a new semi-empirical formula to estimate the 1p-decay half-lives. The robustness of this formula is demonstrated by the accurate predictions of the measured values while making it reliable for forecasting the properties of other potential proton emitters. The phenomenon of shape coexistence as observed in several proton emitters and their respective daughter nuclei, is particularly signicant in this context due to secondary minima in the potential energy surfaces of both the nuclei. Accounting for these factors signicantly affects the estimation of half-lives and branching ratios by introducing additional decay pathways and altering transition probabilities between different nuclear shapes.

## Introduction

Significant advancements in measuring proton decay half-lives through the development of sophisticated facilities like ISOLDE [1], Argonne National Laboratory (ANL) [2], University of Jyväskylä [3], Oak Ridge National Laboratory (ORNL) [4], ISAC-TRIUMF [5] and NSCL [6, 7], and identification of a large number of ground-state and isomeric excited-state proton emitters [8, 9, 10] near the proton drip line has attracted a lot of attention as it can be used as a powerful tool to probe the structure of proton unbound Nilsson orbitals, deformations and decay mechanisms. Precise knowledge of the nuclear structure, lifetimes and disintegration rates of different decay modes, can be useful in various practical applications like selecting isotopes for medical imaging, designing nuclear reactors, dating ancient materials and also in understanding the nucleosynthesis processes in stars [11, 12]. One-Proton (1p) radioactivity is one of the rare decay modes that competes with the $\alpha$ and $\beta$ decay modes near the proton drip lines [13, 14] where the last valence proton remains no longer bound by the strong nuclear forces. Nuclear deformation, that increases the number of nucleons in the classically forbidden regions below the continuum threshold, influences the half-life of decay by altering the potential barrier for particle emission, modifying nuclear matrix elements, and changing the energy levels and transition probabilities [15, 16]. Moreover, the measurements of lifetimes of deformed proton emitters [17, 18] provide information on the last occupied Nilsson configuration and hence the shape of the nucleus which is a fundamental property of the atomic nucleus. This points to the correlation between the nuclear deformations, shape and the life times of deformed proton emitters at the limits of nuclear stability [19, 20, 21] which is one of the most exciting subjects in the contemporary nuclear physics and needs investigation which is the goal of present work.

So far a significant number of proton emitting ground and isomeric states have been identified and investigated in almost all odd Z systems from Sb (Z=51) to Bi (Z=83) with the resulting data becoming a versatile spectroscopic tool [2, 22] to characterize states located near the Fermi surface in nuclei at the threshold of stability. The predicted proton emitters in the region with 50<Z<67 are expected to be quite deformed. Although proton radioactivity provides valuable insights into the nuclear masses, shell structure, and the interactions between bound and unbound nuclear states [16] at the edge of stability, it requires extremely sensitive detection techniques to carefully distinguish these proton decay events from background noise, other radiation types and cosmic rays. However, despite the challenges to produce and isolate the rare isotopes that exhibit proton radioactivity, ongoing improvements in detector technology and collaborations are enhancing the precision to measure the half-life of proton radioactivity [23]. Starting with the first experimentally identified 1p-emitter from the high-spin isomeric state of $^{53}$Co$^m$ in 1970, approximately 52 proton emitters [24] have been discovered so far, decaying from both ground and isomeric states of nuclei [9].

On the theoretical front, the determination of decay half-lives can be done by employing the theoretical models like the distorted-wave Born approximation (DWBA) [25], the two-potential approach (TPA) [25], the quasi-classical method [25], the density-dependent M3Y (DDM3Y) effective interaction [26], the phenomenological unified fission model (PUFM) [27], the RMF+BCS approach [28], the generalized liquid drop model (GLDM) [29], the cluster model [30], the unified fission model [31], the single-folding model (SFM) [32], relativistic density functional theory (RDFT) [33], the method of coupled channel calculations (MCCC) [34], covariant density functional (CDF) theory [35], the deformed density-dependent model [36], the Gamow-like model (GLM) [37], the Coulomb and proximity potential model for deformed nuclei (CPPMDN) [38], the Coulomb and proximity potential model (CPPM) [21], the semiclassical WKB method [39], the non adiabatic quasi particle method [40], and various proximity potentials [14].

In addition, one may also use various (semi) empirical or analytical formulas to determine proton decay half-lives like the formulas proposed by Delion *et al.* [41] or Medeiros *et al.* [42] for both the ground state and isomeric proton transitions. Formulas by Viola-Seaborg and Royer [29] used for $\alpha$-decay, and several other empirical formulas [43, 44, 45, 46, 47] are also used for calculating 1p-decay half-lives.

The existing semi-empirical formulas for proton decay primarily rely on a linear relationship between $log_{10}T_{1/2}$ and $\frac{Z_d}{\sqrt{Q}}$, similar to the approach used for $\alpha$-decay [42, 43, 47]. Previous studies have demonstrated that including deformation effects significantly improves the predictive accuracy of $\alpha$-emission half-lives [48], $\beta$-decay [49] and 2p-emission [50, 51, 52] as well. Interestingly, the studies have indicated that the half-life sensitivity to $Q$-value differs between the proton emission and $\alpha$-decay, owing to a smaller reduced mass and a higher centrifugal barrier in the former case. This difference makes deformation effects particularly important for accurately predicting proton emission half-lives, especially in drip-line nuclei where determining precise angular momentum ($l$) and $Q$-values is challenging.

So far, only a few empirical formulas for 1p-emission consider the quadrupole (and occasionally hexadecapole) deformation of the parent nucleus [51, 53] or the daughter nucleus [47, 54]. However, to the best of our knowledge, no empirical formula has considered the deformation of both the parent and daughter nuclei simultaneously for 1p/2p-emission or $\alpha$-emission. The deformation of the parent nucleus directly affects the potential barrier for particle emission, influencing the tunneling ease, while the deformation of the daughter nucleus affects the final energy state and transition probabilities. Neglecting either aspect can result in significant inaccuracies, as the shape and energy distribution of both the nuclei determine the decay dynamics. By including terms for both parent and daughter deformations, empirical formulas can more accurately capture the complexities of nuclear structure and decay pathways, leading to predictions that align more closely with experimental data and enhancing the formula's reliability and general applicability across the various decay modes and nuclear configurations.

In view of this, we use deformation dependence of proton-emission half-lives for proton-emitters with Z<82 using a semi-empirical formula motivated by the formula used for 2p-radioactivity in our earlier work [52]. The formula includes the quadrupole deformation of both the parent and daughter nuclei in a phenomenological manner in addition to the usual $Q$-value and angular momentum dependence. Our formula proves to be robust that performes well as compared to the other descriptions of proton radioactivity half-lives and correlates the half-lives with nuclear shapes of parent and daughter nuclei. Our calculations reveal that in some shape-coexisting nuclei, the half-lives for transitions from the second minima align more closely with the experimental data. Also, the shape coexistence can alter the branching ratios of nuclear decays by introducing additional decay paths, thereby modifying energy levels and transition rates. These findings underscore the significant impact of shape coexistence on nuclear stability and half-lives, a key outcome of this comprehensive study.

## Theoretical Framework

To calculate the disintegration energy for various decay modes, including proton emission, in unknown nuclei as well as nuclei known to us to some extent and to examine the shapes of all considered nuclei, we utilize two well-established theoretical models: (i) the Macroscopic-Microscopic approach using the Nilsson-Strutinsky method (NSM), and (ii) Relativistic Mean-Field (RMF) theory. A brief overview of these theories is provided below:

### Macroscopic - Microscopic approach using Nilsson-Strutinsky method (NSM)

The Macroscopic-Microscopic approach using the NSM employs a triaxially deformed Nilsson Hamiltonian and the Strutinsky shell correction to study the structural properties of atomic nuclei. This method addresses the interplay between microscopic shell effects and the macroscopic bulk properties of nuclear matter. The well-known Strutinsky shell correction to the energy is obtained by

$$\delta E_{Shell} = \sum_{i=1}^{A} \varepsilon_i - \tilde{E} \tag{1}$$

where $\varepsilon_i$ is the shell model discrete energy and $\tilde{E}$ is the smoothened energy. Hermite polynomials $H_k(u_i)$ upto higher order of correction ensure the smoothened levels. The smearing width of $1.2\hbar\omega$ has been used. The single particle energies $\varepsilon_i$ as a function of deformation parameters ($\beta$, $\gamma$) are generated by Nilsson Hamiltonian for the triaxially deformed oscillator diagonalized in a cylindrical representation [55, 56].

Strutinsky's shell correction $\delta E_{Shell}$ added to macroscopic energy of the spherical drop $BE_{LDM}$ along with the deformation energy $E_{def}$ obtained from surface and Coulomb effects gives the total energy $BE_{gs}$ as in our earlier works [57, 58] corrected for microscopic effects of the nuclear system (For detailed theoretical formalism, please refer [57, 58])

$$BE_{gs}(Z,N,\beta,\gamma) = BE_{LDM}(Z,N) - E_{def}(Z,N,\beta,\gamma) - \delta E_{shell}(Z,N,\beta,\gamma) \tag{2}$$

Energy $E$ $(=-BE)$ minima is searched for $\beta$ (0 to 0.4 in steps of 0.01) and $\gamma$ (from $-180^o$(oblate) to $-120^o$(prolate) and $-180^o < \gamma < -120^o$ (triaxial)).

$\beta$ and $\gamma$ corresponding to E minima provide the equilibrium deformation and shapes of the nuclei. Our predicted $\beta$ values have shown excellent agreement with the experimental values for a wide range of nuclei [58, 59] confirming the reliability of our calculations. The unbound nuclei for one and two protons (1p and 2p) as well as one and two neutrons (1n and 2n) are identified by their respective separation energies: $S_{1p}$, $S_{2p}$, $S_{1n}$, and $S_{2n}$. These separation energies approach zero as they are calculated from the difference between the ground-state binding energies ($BE_{gs}$) of the parent and daughter nuclei.

**Relativistic Mean-Field theory (RMF)**

In case of Relativistic Mean Field (RMF) theory, the Lagrangian density is used to describe the dynamics of nucleons and mesons. We employ NL3$^*$ parameter [60] set, which is a commonly used parameter set in RMF calculations. These calculations typically include nonlinear terms for the scalar $\sigma$ and vector $\omega$ meson to account for the self-interactions of these mesons. Besides the usual nucleon and meson part of the Lagrangian, the interaction and non-linear part of Lagrangian density is given by:

$$\mathscr{L}_{\text{interaction}} = g_\sigma \bar{\psi}\sigma\psi - g_\omega \bar{\psi}\gamma^\mu \omega_\mu \psi - g_\rho \bar{\psi}\gamma^\mu \vec{\tau}\cdot\vec{\rho}_\mu\psi - e\bar{\psi}\gamma^\mu \frac{1+\tau_3}{2}A_\mu\psi \tag{3}$$

$$\mathscr{L}_{\text{nonlinear}} = -\frac{1}{3}g_2\sigma^3 - \frac{1}{4}g_3\sigma^4 + \frac{1}{4}c_3(\omega_\mu\omega^\mu)^2 \tag{4}$$

Here, $g_2$ and $g_3$ are the self-interaction parameters for the $\sigma$ meson, and $c_3$ is the self-interaction parameter for the $\omega$ meson. The nonlinear terms are crucial for reproducing the empirical properties of nuclear matter and finite nuclei within the RMF framework. The NL3$^*$ parameter [60] set provides specific values for the masses and coupling constants that are used in the Lagrangian density. The corresponding Dirac equations for nucleons and Klein-Gordon equations for mesons, derived using the mean-field approximation, are solved using the expansion method on the widely utilized axially deformed Harmonic-Oscillator basis [61, 62]. Quadrupole-constrained calculations have been performed for all the considered nuclei to obtain their potential energy surfaces (PESs) and determine the corresponding ground-state deformations [61, 63]. For nuclei with an odd number of nucleons, a simple blocking method is used without breaking time-reversal symmetry [64]. For further details of these formulations we refer the reader to Refs. [61, 62, 65]

**Half-life calculations**

We perform half-life calculations for the spontaneous emission of protons from a nuclear state relatively long-lived with a lifetime enough to study the atomic structure. We employ a semi-empirical formula, somewhat similar to that used for 2p-radioactivity [52] and incorporate the nuclear deformation of the parent proton emitter and the daughter nuclei apart from the usual Q-value and angular momentum dependence. Although the proton emitting states are unbound to the strong nuclear forces, the structural effects along with Coulomb and centrifugal barriers together can slow down the decay process to provide enough life time for probe. The deformation effects alter the potential barrier, that, the decaying particles must overcome, thereby altering the tunneling probability and thus modifying the half-life. It also impacts the energy levels and transition probabilities which are crucial for determining the decay rates. Hence by including the deformation term, one can ensure the empirical formula to better account for the above factors, leading to more accurate predictions of half-lives which align more closely with the experimental data and thus enhancing its reliability and applicability in predicting the nuclear decay properties.

Including deformation of both the parent and daughter nuclei becomes particularly more important when we are dealing with the nuclei exhibiting shape coexistence which is characterized by one or more nuclear states lying close to the ground state with competing spherical, prolate and oblate shapes having different deformations with very close or similar energies.

The crucial role of interplay between the shell effects and the collective degrees of freedom in the structure and stability of nuclei can be seen in the phenomena of shape coexistence. Such nuclei with coexisting eigenstates, exist in a second minima state lying close to the ground state that may have some finite lifetime sufficient for probes which depends on its excitation energy and the extent of overlap between its wave function and that of the ground state. The existence of two minima in both parent and daughter nuclei results in a possibility that either the parent or daughter nucleus may transit between different shapes during the decay process which may lead to different pathways and different lifetimes for emission.

In a recent work, we reported the deformation dependence of 2p-radioactivity half-lives by including the quadrupole deformation of parent nucleus into an empirical formula which has been able to estimate measured values with high accuracy [52]. Building on this, we use a formula guided by the same principles, selecting the first three terms based on a semi-classical treatment for evaluating the transmission coefficient of an $\alpha$-particle through a Coulomb barrier [66, 67]. The fourth term, $\sqrt{l(l+1)}$, accounts for the hindrance effect of the centrifugal barrier [68]. The deformation term can either take the form of $|\beta|^p$, as chosen phenomenologically in previous works [47, 52, 53, 54], or $(\kappa\beta)^{1/2}\frac{Z}{\sqrt{Q}}$ (where $\beta$ is the deformation of daughter nucleus, and the $\kappa$ is 2 or -1 for prolate or oblate nucleus, respectively), as used recently in a empirical relationship for $\alpha$-decay half-lives [48]. We have compiled a dataset of 52 experimentally identified 1p-emitters, including 37 ground states and 15 isomeric transitions [69]. Given that the nuclei in this study are at or beyond the proton drip line, experimental values of deformation $\beta$ for many of these nuclei are unavailable. Therefore, for fitting purposes, we use theoretical $\beta$ values calculated for all the considered nuclei using the Relativistic Mean Field (RMF) approach [65, 70] and the Macroscopic-Microscopic approach [57, 58]. These calculations were compared with deformation values from the Weizsäcker-Skyrme (WS4) model [71], Hartree-Fock-Bogoliubov (HFB) [72], and Finite Range Droplet Model (FRDM) mass tables [73], among others. We found that the most of the shapes of parent and daughter nuclei were consistent across these theories. However, for the purpose of the best formula fitting, we preferred the deformation values computed by the RMF approach, as done in our previous work [52]. The validity of the formula is evaluated using the k-fold cross-validation technique in machine learning.

Initially, we fitted the term $|\beta|^p$ to the data by varying the power $p$ from 0 to 5 in 0.5 increments, using the deformation of the parent nuclei. The same procedure was repeated using the deformation of the daughter nuclei, and then with both the parent and daughter nuclei. Subsequently, we incorporated the new term $(\kappa\beta)^p \frac{Z}{\sqrt{Q}}$, considering the deformation of (i) the parent nucleus, (ii) the daughter nucleus, and (iii) both nuclei, while varying the power $p$ in each case. Our findings indicate that the new term, accounting for both parent and daughter nuclei deformations, yielded the lowest root mean square error (RMSE) for the data set which highlights the importance of considering the deformation of both parent and daughter nuclei in predicting half-lives. Thus, the final form of the semi-empirical relation proposed by us in this work for estimating the half-life of one-proton decay is as follows:

$$log_{10}T_{1/2} = a + b\sqrt{\mu}\sqrt{Z_d A^{1/3}} + c\sqrt{\mu}\left(\frac{Z_d}{\sqrt{Q}}\right) + d\sqrt{l(l+1)} + e[(-1)^{\kappa_p + \kappa_p\kappa_d}(\kappa_p\beta_p)^{1/2} + (-1)^{\kappa_d + \kappa_p\kappa_d}(\kappa_d\beta_d)^{1/2}]\frac{Z_d}{\sqrt{Q}} \quad (5)$$

In this formula, the half-life is expressed in seconds. The symbols $Z_d$ and $A$ represent the proton number of the daughter nucleus and the mass number of the parent nucleus, respectively. The term $\sqrt{l(l+1)}$, associated with the hindrance effect of the centrifugal barrier, involves the orbital angular momentum $l$ which can be determined using standard selection rules [74]. In the last two term of the formula, the subscript $p$ refers to the parent nucleus, while the subscript $d$ pertains to the daughter nucleus. The parameter $\kappa$ takes the value 2 for prolate nuclei and -1 for oblate nuclei, as modeled in Ref. [48]. This choice reflects the fact that the deformation of the daughter nucleus influences the radius-related factor, which is twice as large for prolate nuclei compared to oblate ones. Additionally, in several decays, the predominance of the prolate shapes has been observed and speculated [52, 58, 75, 76, 77, 78] over the oblate shapes. Therefore, the decays from the prolate parent nuclei are generally more likely than those from oblate parent nuclei. It has also been demonstrated [59] that the transitions between similar shapes like prolate to prolate or oblate to oblate are more probable, while the change in the deformed state from prolate to oblate or oblate to prolate during the decay are less likely (given the available energy). The $\kappa$ term $(-1)^{\kappa_{d(p)} + \kappa_p\kappa_d}$ ensures that the contribution from both terms is summed up when the parent and daughter nuclei have the same deformation state, hence amplifying the effect for prolate-to-prolate transitions and, to a lesser extent, for oblate-to-oblate transitions. The contribution of this term in the half-life estimate will be reduced if the decay involves different deformation states in the parent and daughter nuclei.

The above mentioned formula is fitted through a least squares fit to a total of 52 data points, out of which 34 are the true experimental data and remaining 18 data are listed with their upper limit of half-lives [69]. To incorporate this kind of data, we use the data censoring approach where the actual value is not known precisely but is known to lie below a specified threshold. We have employed Maximum Likelihood Estimation (MLE) to handle the censored data, modifying the likelihood function to account for the probability that the true value is less than or equal to the observed upper limit. Specifically, we have modified the loss function to work alongside the original Mean Squared Error (MSE) loss. For the 34 data points with accurate targets, we continue to use the MSE loss as (target−prediction)$^2$. For the 18 censored data points, we introduced a penalty term that applies only when the model's prediction exceeds the upper bound. In such cases, a penalty of (upper bound−prediction)$^2$ is added to the overall loss. When the prediction is below the upper bound, no penalty is applied. This adjustment ensures that the model accounts for the censored nature of the data without introducing undue bias. We believe this approach better aligns with the nature of the problem and will improve the robustness of our model. It is also important to point out here that out of the 34 true data points, 5 are evaluated using the TNN approach rather than direct measurements. When we exclude these data points and refit the formula, the RMSE and resulting coefficients show no significant change. Therefore, we have included these 5 data points to enhance the accuracy of the predicted half-lives in regions where direct measurements are not available and as has been done in the several Refs. [20, 29, 44, 45, 46, 53, 54, 79].

Finally, fitting the model to all 52 data points yields the following coefficient values: $a=-9.5853$, $b=-0.0516$, $c=0.1135$, $d=0.0620$, and $e=-0.0058$. The positive value of the coefficient $d$ suggests a hindrance effect due to the centrifugal barrier, which can delay the transition and thereby increase the half-life. In contrast, the negative value of the coefficient $e$ indicates a shorter half-life and thus a higher transition probability for deformed nuclei compared to spherical ones. This behavior has also been observed in the contexts of $\alpha$-decay [48] and 2p-decay [52]. To summarize, our proposed formula represented by Eqn. (5) is constructed phenomenologically to predict 1p-decay half-lives, drawing parallels with the physics of $\alpha$-decay and 2p-decay, while incorporating the deformations of both the parent and daughter nuclei, and their respective deformed shapes as well for further improvisation. Hence, this formula is expected to provide much improved accuracy, broader applicability, and better concordance with experimental data, making it a valuable tool for predicting nuclear decay half-lives.

## Results and Discussion

First, we present the fitting of our proposed formula (Eqn. (5)) using the available experimental data of 52 nuclei (37 ground states and 15 isomeric transitions) spanning the region $21 \leq A \leq 185$ along with the experimental $Q$-values and $\beta$ values deduced from RMF theory. Fig. 1 presents RMSE for 34 true experimental data plus 18 upper limit data sets, where we first evaluate the validity of the formula using k-fold cross-validation technique in machine learning which is robust and widely used to evaluate a model's performance and enhance its generalization capabilities [80]. Recently this technique has been successfully applied to assess the formulas for $\alpha$-decay half-life and $\alpha$-particle preformation factors [81] as well. By partitioning the dataset into k equally sized folds ensures that the each data point is used for both training and validation (Each training and test set consist of 80% and 20% data split), thus maximizing the utilization of the available data. During each of the k iterations, the model is trained on (k-1) folds and validated on the remaining fold, with the performance metrics averaged across all iterations to provide a reliable estimate. This approach minimizes the risk of over-fitting and bias associated with single train-test splits, making it essential for model selection and hyper-parameter tuning. Additionally, k-fold cross-validation can adapt to various data structures, including those with imbalanced classes, by employing stratified sampling techniques. It is a crucial tool for enhancing model reliability and ensuring robust evaluation [82, 83].

$$\text{RMSE} = \sqrt{\frac{1}{N_d}\sum_{i=1}^{N_d}(logT_{Th}^i - logT_{Expt.}^i)^2} \quad (6)$$

$$\chi^2 = \frac{1}{N_d - N_p}\sum_{i=1}^{N_d}(logT_{Th}^i - logT_{Expt.}^i)^2 \quad (7)$$

RMSE values shown in Fig. 1 have been computed using Eqn. (6) ($N_d$ is the number of data points). The data in each set is split into 80% training and 20% testing, ensuring that no data is repeated in both training and testing, and that the test data is always different from the previous fold. The RMSE values for both training and testing data across all folds are shown in Fig. 1, with the average values across all folds represented by the final bars. The consistent performance across all folds suggests that the current formula effectively captures the underlying patterns in the data and is not overly sensitive to variations in the training and validation sets. This consistency indicates that the formula is likely to perform well on new, unseen data, which is a key objective for any empirical formula model. The excellent performance across each fold provides confidence in the formula's reliability and making it a valuable tool for practical applications. Overall, the consistent success in k-fold cross-validation strongly indicates a high-quality empirical formula.

Table 1 shows the validation of the formula by comparing our calculated RMSE with other existing formulas specifically developed for 1p-emission [29, 43, 44, 47, 51, 79]. Since most available formulas are tailored for heavy mass proton emitters (Z>50), we calculate the RMSE for 27 data points in the heavy mass region, which, along with the $\chi^2$ values (as defined in Eqn. (7), where $N_p$ refers to the total number of parameters used for fitting), are presented in Table 1. The $\chi^2$ statistics is crucial as it indicates a better fit when it converges to a lower value, allowing each formula to be compared on an equal basis in terms of their fitting parameters. It is evident from the Table 1 that the present formula shows the lowest RMSE and $\chi^2$ values than all the other formulas compared, demonstrating exceptional predictive accuracy and an excellent fit to the data with minimal error. Moreover, the low $\chi^2$ value indicates that the formula's complexity is well-balanced with all the fitting parameters, avoiding over-fitting while accurately modeling the data. This balance makes the formula not only reliable and robust for the current dataset but also likely to generalize well to new and unseen data.

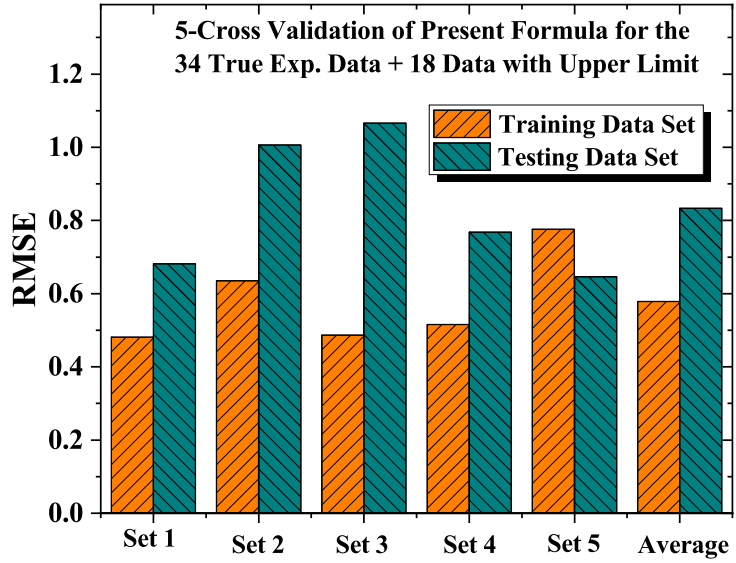

**Figure 1.** RMSE for various data-sets of 34 true experimental data + 18 data with upper limit (37 ground states and 15 isomeric transitions) [69]. Each training and test set consist of 80% and 20% data split. The last bars are showing average of 5 data sets.

**Table 1.** RMSE for 27 true experimental proton decay data for various available formulas of 1p-decay.

| Theory/method | RMSE | $\chi^2$ |
|---|---|---|
| Present Formula | 0.6998 | 0.6010 |
| Yusofvand [79] | 0.8854 | 1.1990 |
| Sreeja [44] | 0.9611 | 1.0917 |
| Chen [46] | 0.9851 | 1.0514 |
| UDL [43] | 1.0008 | 1.1758 |
| Budaca [51] | 1.1577 | 1.8092 |
| Dong1 [29] | 1.3036 | 1.9950 |
| Dong2 [29] | 1.3526 | 2.1476 |
| Soylu [47] | 2.3832 | 7.3833 |

Table 2 lists our calculated half-lives of various heavy proton emitters along with the measured values of $log_{10}T_{1/2}$, and $Q_p$, and $l$ used in establishing the formula in Eqn. (5). The orbital angular momentum quantum number $l$ is calculated based on selection rules derived from the parity and spin of the parent and daughter nuclei, as provided by NUBASE2020 [69]. Due to absence of experimental deformation measurements along the driplines, the deformation values $\beta_p$ and $\beta_d$ for these proton emitters are taken from the RMF theory, as mentioned in the formalism section. The present formula consistently provides reasonably accurate predictions compared to various other models and formulas [20, 26, 27, 28, 29, 36, 38, 41, 42, 44, 45, 46, 51, 53, 54, 79, 84, 85, 86, 87] which demonstrates its capability in predicting lifetimes for a wide range of nuclei ($21 \leq A \leq 185$) more effectively than the other existing models or formulas. This comparison highlights the importance of selecting and fine-tuning empirical formulas to ensure optimal performance and robustness, particularly in regions with limited knowledge or lack of experimental data.

We extend our calculation to estimate the 1p-decay half-lives of proton unbound or nearly unbound nuclei with known $Q_p$ values from AME2020 [88] but unknown half-lives. Table 3 shows that our predicted proton decay half-lives align with the experimental range (lower limit $T_{1/2} \sim 10^{-6}$ sec.), suggesting these nuclei to be potential proton emitters that may be identified in near future. Accurate predictions of half-lives for one-proton emitters are crucial for the discovery of new elements and isotopes, aiding in the synthesis of superheavy elements, and enhancing our understanding of nuclear forces and structure which are often challenging to observe experimentally.

**Table 2.** Comparison of our calculated logarithmic half-lives $log_{10}T_{1/2}$ (column 7) of proton emitters (column 1) with the experimental values (column 6) and other theoretical predictions (columns 8-27) [20, 26, 27, 28, 29, 36, 38, 41, 42, 44, 45, 46, 51, 53, 54, 79, 84, 85, 86, 87]. Experimental $Q_p$, $l$ and half-life values are taken from NUBASE2020 [69] and AME2020 [88]. $\beta$ values computed using RMF theory. Other theories including UDLP(universal decay law for proton decay); GLDM and ELDM (generalized and effective liquid drop models); DDDM (deformed density-dependent model); RMF+BCS (Relativistic mean field theory combined with BCS method); PCM+UFM (preformed cluster model modified into a unified fission model); DDM3Y (density dependent M3Y interaction); JLM (Jeukenne, Lejeune and Mahaux interaction); CPPM (Coulomb and proximity potential model) and CPPMDN (CPPM for the deformed nuclei) have been used for comparison. # values are the estimated from Trends in Neighboring Nuclei (TNN) in NUBASE2020 [69]

| Proton Emitter | $Q_p$ (MeV) | $l$ | $\beta_p$ | $\beta_d$ | Exp. [69] | Present Formula | Sreeja [44] | Delion [41] | UDLP [20] | Zhang [45] | Chen [46] | Budaca [51] | Yusofvand [79] | CPPMDN [38] | DDDM [36] | RMF+BCS [84] | PCM+UFM [27] | DDM3Y [26] | GLDM [29] | JLM [28] | Medeiros [42] | Ni [54] | Dehghani [53] | ELDM [85] | CPPM [86] | Gamow-Like [87] |
|---|---|---|---|---|---|---|---|---|---|---|---|---|---|---|---|---|---|---|---|---|---|---|---|---|---|---|
| | | | | | | | | | | | Ground to Ground Transition | | | | | | | | | | | | | | | |
| $^{109}$I | 0.83±0.004 | 2 | 0.16 | 0.12 | -3.99±0.002 | -4.14 | -3.86 | -4.99 | -3.61 | - | -3.51 | -3.95 | -3.44 | -3.18 | -4.42 | -4.43 | - | - | - | - | -3.98 | -3.82 | -3.88 | -3.91 | -4.28 | - |
| $^{112}$Cs | 0.82±0.004 | 2 | 0.20 | 0.15 | $-3.30^{+0.079}_{-0.097}$ | -3.95 | -3.08 | -4.29 | -2.93 | - | -2.84 | -3.24 | -2.67 | -2.35 | -3.64 | -3.33 | - | - | - | - | -3.28 | -3.00 | -3.10 | -3.13 | -3.57 | - |
| $^{113}$Cs | 0.98±0.002 | 2 | 0.22 | 0.15 | $-4.78^{+0.018}_{-0.019}$ | -4.48 | -5.03 | -5.75 | -4.81 | - | -4.80 | -5.22 | -4.92 | -4.51 | -5.21 | -5.43 | - | - | - | - | -4.87 | -4.95 | -5.06 | -5.30 | -5.55 | - |
| $^{130}$Eu | 1.03±0.200 | 2 | 0.38 | 0.38 | $-3.05^{+0.189}_{-0.169}$ | -4.05 | -3.09 | -3.94 | -3.32 | - | -3.12 | -2.74 | -3.27 | -2.66 | -3.39 | -3.82 | - | - | - | - | -2.82 | -2.59 | -2.65 | -3.49 | -3.76 | - |
| $^{135}$Tb | 1.19±0.007 | 3 | 0.36 | 0.37 | $-3.03^{+0.131}_{-0.116}$ | -4.25 | -3.50 | -4.77 | -3.74 | - | -3.38 | -3.21 | - | -3.55 | -3.75 | -3.81 | - | - | - | - | -2.77 | -3.00 | -3.22 | -3.91 | -3.98 | - |
| $^{141}$Ho | 1.19±0.007 | 3 | 0.33 | 0.21 | $-2.39^{+0.095}_{-0.121}$ | -4.00 | -2.94 | -4.09 | -3.19 | - | -2.81 | -2.57 | - | -2.88 | -2.89 | -1.73 | - | - | - | - | -2.20 | -2.86 | -3.09 | -3.34 | -3.40 | - |
| $^{144}$Tm | 1.73±0.016 | 5 | 0.24 | 0.21 | $-5.57^{+0.002}_{-0.001}$ | -4.76 | -5.39 | - | -4.66 | - | -5.22 | - | -5.35 | -6.12 | -5.20 | -5.72 | -5.14 | -5.17 | - | - | - | - | -4.79 | -5.14 | -4.97 | - |
| $^{145}$Tm | 1.75±0.007 | 5 | 0.33 | 0.20 | -5.41±0.027 | -4.84 | -5.57 | -9.50 | -4.83 | -5.41 | -5.40 | -4.92 | -5.52 | -6.32 | -5.20 | -5.72 | -5.14 | -5.17 | -5.66 | -5.10 | -5.24 | -5.33 | -5.09 | -5.42 | -5.16 | - |
| $^{155}$Ta | 1.45±0.015 | 5 | 0.03 | 0.06 | $-2.54^{+0.225}_{-0.165}$ | -3.71 | -2.32 | -8.86 | -2.25 | -2.30 | -2.40 | -2.21 | -2.48 | -3.24 | -2.31 | -2.45 | -4.65 | -4.65 | -2.41 | -4.57 | -4.57 | -4.59 | -2.57 | -2.32 | -2.27 | - |
| $^{159}$Re | 1.82±0.020 | 5 | 0.05 | 0.09 | # $-4.68^{+0.076}_{-0.092}$ | -4.30 | -4.46 | - | - | -4.59 | -4.49 | -4.39 | - | -5.42 | -4.40 | -4.67 | - | - | -4.64 | - | -4.31 | - | -4.64 | -4.64 | -4.43 | - |
| $^{160}$Re | 1.28±0.007 | 2 | 0.11 | 0.10 | $-3.05^{+0.079}_{-0.048}$ | -3.42 | -2.35 | -3.84 | -2.91 | -3.15 | -2.45 | -3.05 | -2.88 | -1.99 | -3.17 | -3.14 | -3.00 | -3.11 | -3.11 | -2.87 | -3.48 | -3.16 | -3.08 | -2.92 | -3.09 | -3.48 |
| $^{161}$Re | 1.21±0.005 | 0 | 0.13 | 0.11 | -3.43±0.010 | -3.39 | -3.18 | -3.43 | -2.84 | -3.38 | -3.28 | -3.05 | -2.96 | -1.57 | -3.44 | -3.35 | -3.46 | -3.23 | -3.32 | -3.29 | -3.43 | - | -2.99 | -2.95 | -3.15 | -3.64 |
| $^{164}$Ir | 1.84±0.009 | 5 | 0.11 | 0.11 | # $-3.96^{+0.190}_{-0.139}$ | -4.27 | -4.14 | -7.77 | -4.08 | -4.15 | -4.25 | - | -3.93 | -5.17 | -4.04 | -4.24 | -3.92 | -4.19 | -4.21 | -3.86 | - | -3.93 | -4.46 | -4.38 | -4.21 | - |
| $^{170}$Au | 1.49±0.012 | 2 | 0.04 | 0.12 | $-3.49^{+0.075}_{-0.060}$ | -3.56 | -3.11 | - | -3.82 | - | -3.25 | -4.03 | -3.86 | -2.95 | - | - | - | - | - | - | -3.08 | - | -4.04 | -3.93 | -3.98 | - |
| $^{171}$Au | 1.47±0.010 | 0 | -0.01 | 0.12 | $-4.65^{+0.185}_{-0.151}$ | -3.51 | -4.39 | -4.77 | -4.26 | -4.92 | -4.46 | -4.54 | -4.18 | -3.13 | -4.93 | -4.88 | -5.02 | -4.79 | -4.87 | -4.86 | -3.74 | -4.40 | -4.47 | -4.53 | -4.57 | -3.22 |
| $^{176}$Tl | 1.28±0.018 | 0 | 0.01 | -0.09 | $-2.21^{+0.025}_{-0.012}$ | -2.62 | -2.35 | - | -2.07 | - | -2.36 | -2.25 | -2.18 | -0.59 | - | - | - | - | - | - | -2.72 | - | -2.16 | -1.91 | -2.13 | - |
| $^{177}$Tl | 1.18±0.019 | 0 | 0.00 | -0.12 | -1.17±0.121 | -2.19 | -1.33 | -1.05 | -0.92 | -1.12 | -1.26 | -4.33 | -1.15 | 0.71 | -0.91 | -1.05 | -1.36 | -0.99 | -1.05 | -1.20 | -1.14 | -0.74 | -0.96 | -0.61 | -0.86 | -5.00 |
| $^{185}$Bi | 1.62±0.080 | 0 | -0.17 | -0.06 | # $-4.23^{+0.191}_{-0.349}$ | -3.66 | - | - | -4.72 | - | - | -4.23 | - | - | - | - | - | - | -3.39 | -5.36 | - | -4.47 | - | - | - | - |
| | | | | | | | | | | | Isomer to Ground Transition | | | | | | | | | | | | | | | |
| $^{141}$Ho$^m$ | 1.26±0.007 | 0 | 0.33 | 0.21 | $-5.18^{+0.196}_{-0.197}$ | -4.38 | -5.58 | -5.18 | -5.22 | -5.58 | -5.78 | -4.86 | -5.30 | -4.44 | -5.83 | -4.88 | - | - | - | - | -4.51 | -5.19 | -5.79 | - | -5.77 | -5.87 |
| $^{147}$Tm$^m$ | 1.14±0.006 | 2 | -0.19 | -0.16 | -3.44±0.048 | -3.51 | -2.54 | -4.28 | -2.82 | 3.43 | 0.68 | -3.04 | -2.72 | -1.99 | -3.37 | -3.43 | -3.39 | -3.20 | -3.44 | -3.27 | -3.77 | -3.63 | -3.32 | - | 0.61 | 0.87 |
| $^{150}$Lu$^m$ | 1.32±0.033 | 2 | -0.17 | -0.16 | $-4.52^{+0.007}_{-0.005}$ | -3.81 | -3.71 | -5.34 | -4.03 | -4.76 | -3.63 | -4.27 | -4.05 | -3.33 | -4.71 | -4.81 | -4.38 | -4.56 | -4.76 | -4.24 | -4.96 | -4.77 | -4.20 | - | -4.73 | -4.37 |
| $^{151}$Lu$^m$ | 1.33±0.002 | 2 | -0.14 | -0.15 | -4.80±0.027 | -3.83 | -3.84 | -5.61 | -4.27 | -4.92 | -3.90 | - | -4.35 | -3.61 | -4.86 | -4.88 | -4.91 | -4.73 | -4.96 | -4.91 | -4.96 | - | -4.55 | - | -4.58 | -4.66 |
| $^{159}$Re$^m$ | 1.83±0.020 | 5 | 0.05 | 0.09 | -4.70±0.087 | -4.33 | -4.55 | - | -4.32 | -4.55 | -4.59 | - | -4.38 | -5.52 | -4.49 | -4.71 | -0.60 | -0.46 | - | - | - | - | -4.43 | - | -4.74 | -4.52 |
| $^{164}$Ir$^m$ | 1.85±0.009 | 5 | 0.11 | 0.11 | -3.96±0.190 | -4.41 | - | - | - | - | - | - | - | - | - | - | - | - | - | - | -3.77 | - | - | - | - | - |
| $^{165}$Ir$^m$ | 1.73±0.007 | 5 | 0.10 | 0.12 | -3.4±0.087 | -3.98 | -3.42 | - | -3.41 | -3.39 | -3.55 | -3.43 | -3.20 | -4.42 | -3.29 | -3.48 | -3.51 | -3.43 | -3.46 | -3.44 | -3.46 | - | -3.41 | -4.17 | -3.69 | -3.50 |
| $^{170}$Au$^m$ | 1.77±0.012 | 5 | 0.04 | 0.12 | $-2.98^{+0.035}_{-0.028}$ | -3.95 | -3.17 | - | -3.32 | -3.17 | -3.33 | -3.31 | -2.91 | -4.24 | -2.33 | -2.63 | -3.01 | -2.20 | - | - | -2.77 | - | -3.01 | - | -3.44 | -3.31 |
| $^{171}$Au$^m$ | 1.72±0.010 | 5 | -0.01 | 0.12 | -2.65±0.013 | -3.71 | -2.81 | -6.41 | -3.00 | -2.60 | -2.99 | -2.99 | -2.56 | -3.87 | -2.33 | -2.63 | -3.03 | -2.20 | -2.61 | -2.96 | -2.46 | -2.61 | -2.64 | - | -3.14 | -2.97 |
| $^{177}$Tl$^m$ | 1.99±0.019 | 5 | 0.00 | -0.12 | -3.35±0.076 | -3.80 | -4.05 | -7.01 | -4.30 | -3.45 | -3.64 | - | -3.74 | -5.28 | -3.28 | -3.47 | -4.49 | -4.38 | -3.47 | -4.46 | -3.60 | -3.84 | -3.68 | - | -3.31 | -3.03 |
| $^{185}$Bi$^m$ | 1.62±0.080 | 0 | -0.17 | -0.06 | $-4.23^{+0.201}_{-0.105}$ | -3.66 | -4.61 | - | - | -4.61 | -4.73 | - | -4.47 | -3.60 | -4.13 | -3.38 | -5.44 | -5.18 | - | - | -4.36 | - | -4.02 | -4.89 | -5.02 | -4.97 |

**Table 3.** Proton decay half-lives are predicted for new potential candidates of one-proton decay near the proton drip line using data from NUBASE2020 [69] and AME2020 [88].

| Proton Emitter | $Q_p$ (MeV) | $l$ | $\beta_p$ | $\beta_d$ | $log_{10}T_{1/2}$ (sec.) Present Formula | Proton Emitter | $Q_p$ (MeV) | $l$ | $\beta_p$ | $\beta_d$ | $log_{10}T_{1/2}$ (sec.) Present Formula |
|---|---|---|---|---|---|---|---|---|---|---|---|
| $^{24}$P | 2.78 | 2 | -0.14 | -0.21 | -8.86 | $^{52}$Cu | 2.48 | 1 | 0.19 | -0.03 | -8.07 |
| $^{32}$K | 3.38 | 2 | -0.12 | -0.22 | -8.77 | $^{56}$Ga | 3.14 | 1 | 0.26 | 0.25 | -8.28 |
| $^{33}$Ca | 1.75 | 2 | -0.04 | -0.12 | -8.27 | $^{57}$Ga | 2.69 | 1 | 0.24 | 0.24 | -8.14 |
| $^{35}$Sc | 4.92 | 3 | -0.09 | 0.00 | -8.79 | $^{58}$Ga | 1.72 | 3 | 0.24 | 0.19 | -7.52 |
| $^{37}$Sc | 2.94 | 3 | -0.10 | 0.00 | -8.50 | $^{60}$As | 3.44 | 1 | 0.22 | 0.21 | -8.26 |
| $^{39}$V | 3.91 | 3 | 0.28 | 0.22 | -8.66 | $^{61}$As | 3.04 | 1 | -0.19 | 0.13 | -7.91 |
| $^{40}$V | 2.68 | 3 | -0.18 | -0.15 | -8.37 | $^{62}$As | 2.08 | 1 | 0.22 | 0.19 | -7.74 |
| $^{41}$V | 2.02 | 3 | -0.18 | -0.15 | -8.15 | $^{67}$Br | 1.84 | 1 | -0.28 | -0.25 | -7.43 |
| $^{41}$Cr | 0.65 | 1 | 0.23 | -0.18 | -6.80 | $^{71}$Rb | 1.52 | 3 | -0.32 | -0.31 | -6.90 |
| $^{43}$Mn | 3.02 | 3 | -0.18 | -0.17 | -8.36 | $^{79}$Nb | 1.91 | 4 | 0.54 | 0.49 | -7.06 |
| $^{47}$Co | 2.12 | 3 | 0.05 | 0.00 | -7.89 | $^{83}$Tc | 1.76 | 1 | -0.22 | 0.59 | -6.33 |
| $^{48}$Co | 1.57 | 3 | -0.08 | -0.07 | -7.60 | | | | | | |

From the data in Tables 2 and 3, it is evident that the nuclear shape of the parent nucleus is usually preserved in the daughter nucleus following the decay process. For instance, if the parent nucleus is prolate, the daughter nucleus tends to maintain a prolate shape. Similar is the case for the other shapes. This nature of preserving the shape may be attributed to the relatively minor perturbation caused by the emission of a single proton, which does not alter the deformation and overall structure of the nucleus significantly. Consequently, the decay pathways and the corresponding branching ratios are influenced by the initial shape configuration. However, if the deformation and shape changes from emitter to residual nuclei during the decay process, our formula has been fine tuned to incorporate the changes effectively. In the present work, we have rarely found the combination of parent and daughter with significant change in shape.

**Table 4.** Our calculated half-lives of for the potential one proton emitter candidates, estimated using the present formula (Eqn. (5)) where $Q_p$ and $\beta$ are taken from NSM and RMF. The angular momentum ($l$) is calculated by using parity and spin of parent and daughter nuclei, which are taken from NUBASE2020 [69] or from Ref. [89] (see the text for detail).

| Proton Emitter | $l$ | RMF | | | | NSM | | | |
|---|---|---|---|---|---|---|---|---|---|
| | | $Q_p$ (MeV) | $\beta_p$ | $\beta_d$ | $log_{10}T_{1/2}$ (Sec.) | $Q_p$ (MeV) | $\beta_p$ | $\beta_d$ | $log_{10}T_{1/2}$ (Sec.) |
| $^{75}$Zr | 1 | 0.19 | -0.29 | -0.31 | -0.65 | 0.12 | -0.18 | -0.25 | 1.97 |
| $^{83}$Ru | 1 | 0.44 | -0.16 | -0.23 | -3.32 | 0.81 | -0.21 | -0.21 | -5.25 |
| $^{125}$Sm | 1 | 0.47 | 0.39 | 0.38 | -1.31 | 0.45 | 0.32 | 0.32 | -1.03 |
| $^{130}$Gd | 5 | 0.15 | 0.39 | 0.38 | 6.28 | 0.17 | 0.32 | 0.32 | 5.63 |
| $^{144}$Yb | 2 | 0.42 | 0.27 | 0.30 | -0.64 | 0.66 | 0.22 | 0.22 | -1.63 |

Table 4 lists the potential one-proton emitters (with one proton separation energy $S_P \leq 0$) predicted by using our theoretical formalisms i.e. NSM and RMF models. The predicted half-lives and deformations from both the theories are found to be reasonably consistent and within the expected experimental range while exhibiting shape preservation in the decay process. Experimental investigations of these predicted emitters will not only help validate theoretical models but also enhance our understanding of the proton drip line. These studies will contribute to advancements in nuclear astrophysics and particle physics, offering deeper insights into nuclear stability and the fundamental processes governing the structure and behavior of atomic nuclei.

### Correlation between half-life and shape coexistence

The relationship between the half-life of a decay and shape coexistence in parent and daughter nuclei is a complex phenomenon where the nucleus can adopt multiple distinct shapes, such as spherical, prolate, or oblate, at nearly the same energy levels. This structural flexibility can significantly influence the decay process, as transitions between different shapes can modify the potential energy landscape and affect decay pathways, as discussed in Ref. [59]. When both the parent and daughter nuclei exhibit shape coexistence, the half-life of the decay can be notably affected due to variations in nuclear deformation and consequent changes in quantum mechanical tunneling probabilities. These changes can either enhance or impede the decay process, depending on the energy barriers and the degree of overlap between the nuclear wave functions of the initial and final states.

The Nilsson-Strutinsky Method (NSM) has been found very effective [58, 59] in predicting the shape coexistence which shows various $\gamma$ competing for minima and sometimes we witness multiple minima representing different shapes. In the Relativistic Mean-Field (RMF) approach, we determine the nuclear deformation by solving the RMF equations (the Dirac equation for nucleons) in a deformed potential minimizing the total energy [61, 63]. Both the NSM and RMF approaches have been successfully used to reproduce experimental deformation across the nuclear chart [58, 59, 90].

It is important to emphasize that for all the nuclei of interest in this study (as discussed in the previous subsection), we have calculated the potential energy surfaces using both the NSM and RMF theories. The deformation values referenced throughout this article correspond to the nuclear state with the minimum energy. Among the identified 1p-emitters, approximately 30 parent nuclei and a similar number of daughter nuclei exhibit shape coexistence, where two energy minima with nearly identical energies are present. To illustrate this, Fig. 2 shows the potential energy surfaces of the one-proton emitters $^{67}$Kr and $^{71}$Rb, along with their respective daughter nuclei, $^{66}$Br and $^{70}$Kr, as calculated using both the NSM and RMF approaches. The plots reveal multiple energy minima, clearly indicating shape coexistence with only a small energy difference between them. The results from both methods are in close agreement, reinforcing the significant role of nuclear deformation in determining binding energy. These findings suggest that the shape of a nucleus can transition between spherical, prolate, and oblate forms depending on the specific nucleus.

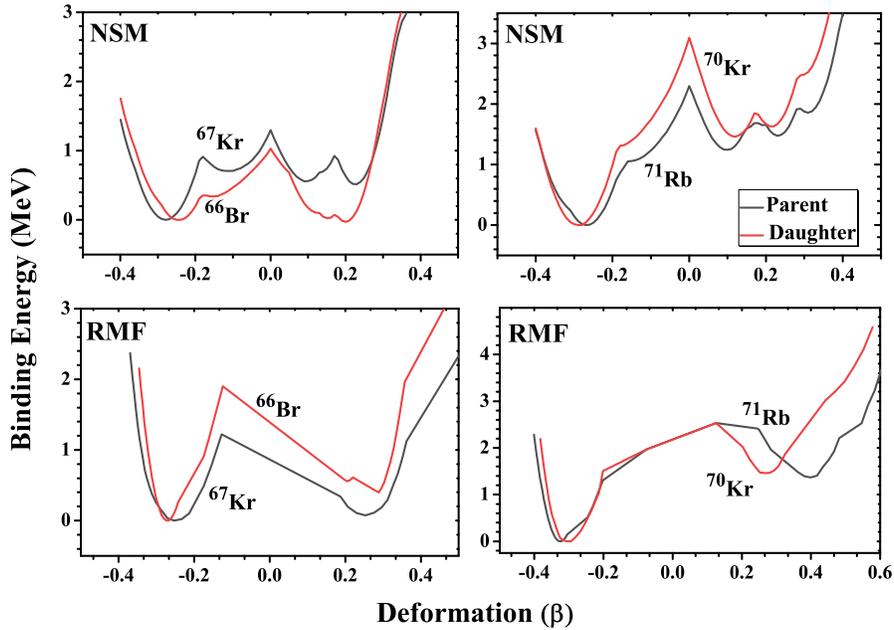

**Figure 2.** Potential energy surfaces (PES) for the one-proton emitters $^{67}$Kr and $^{71}$Rb, along with their corresponding daughter nuclei $^{66}$Br and $^{70}$Kr. The energies are normalized to zero with respect to the lowest energy minima for each nucleus, allowing for a clear comparison of the deformation states and their relative stability.

In order to establish the correlations between half-lives and shape coexistence, we selected only those parent-daughter combinations where both nuclei exhibit shape coexistence, and the energy difference between the two minima is around 1 MeV or less (i.e., $\Delta E \lesssim 1$ MeV). This choice of energy difference between shape minima allows for relatively easy transitions between shapes through thermal excitations, electromagnetic transitions, or other low-energy processes. The near-degeneracy in energy levels enables the nucleus to explore different configurations, resulting in a rich spectrum of states and complex decay pathways. The small energy differences facilitate dynamic shifts from one shape to another during radioactive decay under certain conditions.

Here we examine the various possibilities of the transitions during the decay process for a few selected nuclei, such as: (i) the ground state (the minimum with the least energy) of the parent nucleus decaying to the ground state of the daughter nucleus (G-G), (ii) the ground state of the parent nucleus decaying to the second minimum of the daughter nucleus (G-S), (iii) the parent nucleus being in the second minimum state and decaying from this second minimum to the ground state of the daughter nucleus (S-G), and (iv) the second minimum of the parent nucleus decaying to the second minimum of the daughter nucleus (S-S).

Four of these possible transition states between the parent and daughter nuclei may significantly influence the decay energy ($Q$-values) and deformation state, thereby altering the half-lives and overall stability of the parent nucleus. With this in mind, we calculated the half-lives for all possible transitions (G-G, G-S, S-G, and S-S) for above mentioned selected nuclei. Our calculated half-lives for various transitions show close agreement with experimental values, such as in the cases of $^{113}$Cs for the NSM approach and $^{65}$Br for the RMF approach where the shape coexistence has negligible impact on half-lives. But in some cases, a significant impact on half-lives is seen due to the changes in $Q$-values by several hundred keV and variations in deformation. However, the ground-state to ground-state (G-G) transitions still align with the experimental data in most nuclei and validating our calculations, but the discrepancies in few nuclei suggest the transitions involving ground state to second minima (G-S), second minima to ground state (S-G), and second minima to second minima (S-S) in shape-coexisting nuclei that might have inuenced the lifetimes, particularly when G-G transitions is not aligning well with the experimental data as is seen in Fig. 3. In few nuclei, the S-G, S-S or G-S transitions are matching better with data demonstrating qualitatively how the nuclear shapes affect half-lives. Despite the simplicity of this approach, the results obtained in this preliminary work in this area are meaningful and useful. The present analysis, although premature or may be subject to uncertainties in theoretical calculations, but certainly highlights the importance of considering the interplay between nuclear shapes and energy states in understanding the decay properties and lifetimes of nuclei, particularly those exhibiting shape coexistence.

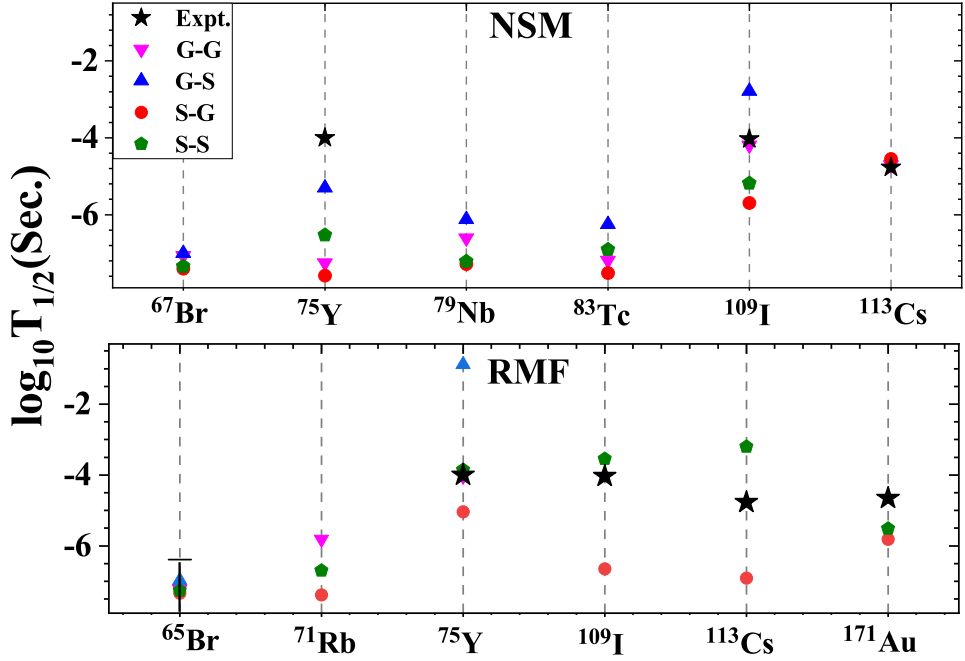

**Figure 3.** 1p-decay half-lives related to various transitions for selected shape coexisting nuclei as found from NSM and RMF calculations. Here, G-G represents transition from ground state of parent nucleus to ground state of daughter nucleus. Similarly the others, where S refers to second minima state of the nucleus (see text for details). We have shown the half-life of $^{65}$Br with upper limit.

Building on the above analysis and recognizing the significant impact of nuclear shapes on half-lives, we can extend this understanding to the branching ratios of various possible decay modes. The likelihood of transitions between different nuclear shapes can vary considerably, with shape coexistence potentially enhancing certain decay channels while suppressing others. This variability directly influences the observed branching ratios, making it essential to consider shape coexistence when predicting and analyzing nuclear decay processes. In this work, our focus is on nuclei near the proton drip-line, where competing decay modes include $\beta^+$-decay and $\alpha$-decay (particularly for $Z > 50$), in addition to 1p-decay. Another rare and exotic decay mode that can occur in certain proton-rich nuclei beyond the proton drip line is 2p-emission, where the simultaneous emission of two protons becomes energetically favorable [52]. The half-lives for these other decay modes $\beta^+$, $\alpha$-decay, and 2p-decay can be estimated using the latest empirical formulas, which have been fitted to the most recent experimental data and found to provide accurate half-life predictions. Specifically, the half-lives for $\beta^+$-decay ($T_{1/2}^{\beta^+}$), $\alpha$-decay ($T_{1/2}^{\alpha}$), and 2p-decay ($T_{1/2}^{2p}$) are calculated using the empirical formulas given in Refs. [91], [81], and [52], respectively. The branching ratios for these respective decay modes can be defined as follows:

$$BR = \frac{T_{1/2}^{Th.}}{T_{1/2}^{1p/\beta^+/\alpha/2p}} \tag{8}$$

where, $T_{1/2}^{Th}$ is the total half-life calculated by considering half-lives of all considered decay modes and the relation is given by:

$$\frac{1}{T_{1/2}^{\text{Th.}}} = \frac{1}{T_{1/2}^{1p}+} + \frac{1}{T_{1/2}^{\beta^+}} + \frac{1}{T_{1/2}^{\alpha}} + \frac{1}{T_{1/2}^{2p}} \tag{9}$$

**Table 5.** Branching ratios related to various decays of a few selected nuclei calculated for G-G, G-S, S-G and S-S transitions by using NSM and RMF approaches. The probable experimental decay modes in second column are taken from NUBASE2020 [69] (see the text for detail).

| Nucleus | Decay Modes [69] | Branching Ratios (BR) | | | | | | | | | | | | | | | |
|---|---|---|---|---|---|---|---|---|---|---|---|---|---|---|---|---|---|
| | | G-G | | | | G-S | | | | S-G | | | | S-S | | | |
| | | p | $\beta^+$ | $\alpha$ | 2p | p | $\beta^+$ | $\alpha$ | 2p | p | $\beta^+$ | $\alpha$ | 2p | p | $\beta^+$ | $\alpha$ | 2p |
| | | | | | | From NSM Approach | | | | | | | | | | | |
| $^{67}$Kr | 2p=0.37; $\beta^+$=? | - | 0.00 | - | 1.00 | - | 0.01 | - | 0.99 | 0.00 | 0.00 | - | 1.00 | 0.00 | 0.00 | - | 1.00 |
| $^{72}$Rb | p? | 1.00 | 0.00 | - | - | - | 1.00 | - | - | 1.00 | 0.00 | - | - | 1.00 | 0.00 | - | - |
| $^{71}$Sr | - | 0.00 | 0.00 | - | 1.00 | - | 1.00 | - | 0.00 | 0.01 | 0.00 | - | 0.99 | - | 0.00 | - | 1.00 |
| $^{115}$Ce | - | 1.00 | 0.00 | 0.00 | 0.00 | 0.17 | 0.83 | 0.00 | 0.00 | 1.00 | 0.00 | 0.00 | 0.00 | 0.90 | 0.10 | 0.00 | 0.00 |
| $^{171}$Ir | $\beta^+$=?; $\alpha$=0.15 | 0.03 | 0.97 | 0.00 | - | - | 1.00 | 0.00 | - | 1.00 | 0.00 | 0.00 | - | - | 1.00 | 0.00 | - |
| $^{170}$Au | p=0.89; $\alpha$=11 | 1.00 | - | 0.00 | - | - | - | 1.00 | - | 1.00 | - | 0.00 | - | 1.00 | - | 0.00 | - |
| $^{175}$Au | $\alpha$=0.88; $\beta^+$=? | 0.90 | 0.10 | 0.00 | - | - | 1.00 | 0.00 | - | - | 1.00 | 0.00 | - | - | 1.00 | 0.00 | - |
| $^{169}$Hg | - | 0.00 | 0.00 | 0.00 | 1.00 | - | 1.00 | 0.00 | 0.00 | 0.00 | 0.00 | 0.00 | 1.00 | 0.00 | 0.00 | 0.00 | 1.00 |
| $^{177}$Tl | $\alpha$=0.73; p? | 1.00 | - | 0.00 | - | - | - | 1.00 | - | - | - | 1.00 | - | - | - | 1.00 | - |
| $^{175}$Pb | - | 0.23 | 0.77 | 0.00 | 0.00 | - | 1.00 | 0.00 | 0.00 | - | 1.00 | 0.00 | 0.00 | - | 1.00 | 0.00 | 0.00 |
| | | | | | | From RMF Approach | | | | | | | | | | | |
| $^{67}$Br | p? | 1.00 | 0.00 | - | - | - | 1.00 | - | - | 1.00 | 0.00 | - | - | 1.00 | 0.00 | - | - |
| $^{68}$Br | p? | - | 1.00 | - | - | - | 1.00 | - | - | 1.00 | 0.00 | - | - | 1.00 | 0.00 | - | - |
| $^{67}$Kr | 2p=0.37; $\beta^+$=? | 1.00 | 0.00 | - | 0.00 | - | 1.00 | - | 0.00 | 1.00 | 0.00 | - | 0.00 | - | 0.53 | - | 0.47 |
| $^{71}$Rb | p? | 1.00 | 0.00 | - | - | - | 1.00 | - | - | 1.00 | 0.00 | - | 0.00 | 1.00 | 0.00 | - | - |
| $^{72}$Rb | p? | 0.04 | 0.96 | - | - | - | 1.00 | - | - | 1.00 | 0.00 | - | - | - | 1.00 | - | - |
| $^{71}$Sr | - | - | 1.00 | - | - | - | 1.00 | - | - | 1.00 | 0.00 | - | - | - | 1.00 | - | - |
| $^{77}$Y | $\beta^+$?; $\beta^+$+p?; p? | - | 1.00 | - | - | - | 1.00 | - | - | 1.00 | 0.00 | - | - | - | 1.00 | - | - |
| $^{79}$Nb | $\beta^+$?; $\beta^+$+p?; p? | 1.00 | 0.00 | - | - | 1.00 | 0.00 | - | - | 1.00 | 0.00 | - | - | - | 1.00 | - | - |
| $^{81}$Nb | $\beta^+$?; $\beta^+$+p?; p? | 0.06 | 0.94 | - | - | - | 1.00 | - | - | 1.00 | 0.00 | - | - | - | 1.00 | - | - |
| $^{83}$Tc | $\beta^+$?; $\beta^+$+p?; p? | 1.00 | 0.00 | - | - | - | 1.00 | - | - | 1.00 | 0.00 | - | - | 1.00 | 0.00 | - | - |
| $^{89}$Rh | $\beta^+$?; $\beta^+$+p?; p? | - | 1.00 | - | - | - | - | - | - | 1.00 | 0.00 | - | - | - | - | - | - |
| $^{170}$Au | p=0.89; $\alpha$=0.11 | 1.00 | - | 0.00 | - | 0.00 | - | 1.00 | - | 1.00 | - | 0.00 | - | 1.00 | - | 0.00 | - |

The half-lives and corresponding branching ratios have been calculated for cases where both the parent and daughter nuclei exhibit shape coexistence, similar to the method used for Fig. 3. While the nuclei in Fig. 3 demonstrate nearly 100% probability of 1p-emission, the selected nuclei here have the potential for multiple competing decay modes, including one-proton emission, $\beta^+$-decay, $\alpha$-decay, and 2p-decay. The likelihood of these decays can vary depending on the specific shape configurations of the parent and daughter nuclei. To illustrate the impact of shape configurations on branching ratios, we have listed the branching ratios for probable decay modes of a few selected nuclei in Table 5. These ratios are calculated for ground-state to ground-state (G-G), ground-state to second minima (G-S), second minima to ground-state (S-G), and second minima to second minima (S-S) transitions, using the NSM and RMF approaches. The second column of the table lists the probable experimental decay modes, sourced from NUBASE2020 [69]. A branching ratio of 0.00 indicates a finite, albeit very small, probability of decay, while a blank cell signifies that the particular decay is energetically forbidden.

An analysis of the Table 5 reveals several intriguing outcomes of this study. For instance, in the case of the $^{67}$Kr nucleus, the recent measurements of its 2p-decay lifetime have demonstrated significant influence of shape and deformation on 2p-radioactivity [92], as corroborated by the data in Table 5. This nucleus meets all the criteria for 2p-emission, resulting in a 37(14)% 2p branching ratio [93]. Our theoretical results derived from both the NSM and RMF approaches confirm that $^{67}$Kr predominantly decays via 2p emission, but also indicate a finite probability of $\beta^+$-emission if the transition occurs from the ground state of $^{67}$Kr to the second minima state of its daughter nucleus. Notably, RMF theory predicts a branching ratio of 47% for 2p-emission and 53% for $\beta^+$ emission if the transition proceeds from the second minima of the parent to the second minima of the daughter. Another example is $^{170}$Au, which, according to NUBASE2020, has an 89% probability of one-proton emission and an 11% probability of $\alpha$-emission. Our theoretical analysis suggests that $\alpha$-emission could occur if the transition is from the ground state to the second minima of its daughter, while the one-proton emission dominates in all the other scenarios.

These interesting observations underscore the impact of nuclear shapes and transitions on decay modes, highlighting the shifting probabilities of various decay pathways depending on the shape states involved. This behavior offers profound insights into nuclear structure, revealing the complex interplay between shape degrees of freedom and underlying nuclear forces. However, it is important to incorporate theoretical uncertainties into decay lifetimes and branching ratios to accurately reflect the confidence and reliability of these predictions. While this study represents a preliminary exploration in this area, it clearly illustrates the significant influence of shape-coexisting states on nuclear structure, stability, and lifetimes. These promising results motivate further research to obtain more detailed and refined insights.

## Conclusions

The impact of nuclear deformations on 1p-decay half-lives has been studied in a theoretical framework for nuclei with Z<82 by considering the deformation of both the parent and daughter nuclei in the decay process. By employing a newly proposed semi-empirical formula that incorporates nuclear deformation phenomenologically, we obtained precise estimates of measured proton decay half-lives and reliably identified new potential proton emitters. The deformation of the parent nucleus influences the potential barrier, while the deformation of the daughter nucleus affects the disintegration energy, hence the accurate half-life predictions are reported by considering the deformed shapes of both the parent and residual nuclei involved in the decay process. Also, we observed shape coexistence in several proton emitters and their daughter nuclei, which is significant due to the presence of secondary minima in the potential energy surfaces of both nuclei. This phenomenon of coexisting states not only impacts the accuracy of half-life estimates but also affects branching ratios by introducing additional decay pathways and altering transition probabilities between different nuclear shapes.

## Data availability

We are committed to supporting open science and data transparency. The authors declare that the experimental data used in the article are available on the National Nuclear Data Center (https://www.nndc.bnl.gov/). The other data supporting the findings of this study are available within the paper in the form of tables. The left datasets used and/or analyzed during the current study are available from the corresponding author upon reasonable request.

## References

1. Catherall, R. *et al.* The isolde facility. *J. Phys. G: Nucl. Part. Phys.* **44**, 094002 (2017).
2. Woods, P. & Davids, C. Nuclei beyond the proton drip-line. *Annu. Rev. Nucl. Part. Sci.* **47**, 541–590 (1997).
3. Saastamoinen, A. *et al.* Experimental study of $\beta$-delayed proton decay of al 23 for nucleosynthesis in novae. *Phys. Rev. CNuclear Phys.* **83**, 045808 (2011).
4. Baktash, C. *et al.* Nuclear physics at oak ridge national laboratory. *Nucl. Phys. News* **12**, 4–12 (2002).
5. Ayyad, Y. *et al.* Direct observation of proton emission in be 11. *Phys. Rev. Lett.* **123**, 082501 (2019).
6. Gillespie, S. A. *et al.* Proton decay spectroscopy of s 28 and cl 30. *Phys. Rev. C* **105**, 044321 (2022).
7. Friedman, M. *et al.* Low-energy al 23 $\beta$-delayed proton decay and na 22 destruction in novae. *Phys. Rev. C* **101**, 052802 (2020).
8. Détraz, C. & Vieira, D. J. Exotic light nuclei. *Annu. Rev. Nucl. Part. Sci.* **39** (1989).
9. Blank, B. & Borge, M. Nuclear structure at the proton drip line: Advances with nuclear decay studies. *Prog. Part. Nucl. Phys.* **60**, 403–483 (2008).
10. Sonzogni, A. Proton radioactivity in z> 50 nuclides. *Nucl. Data Sheets* **95**, 1–48 (2002).
11. Rogers, A. M. *et al.* Ground-state proton decay of br 69 and implications for the se 68 astrophysical rapid proton-capture process waiting point. *Phys. Rev. Lett.* **106**, 252503 (2011).
12. Auranen, K. *et al.* Proton decay of 108i and its significance for the termination of the astrophysical rp-process. *Phys. Lett. B* **792**, 187–192 (2019).
13. Cai, B. *et al.* Isomeric structure in the sn 100 region: Possible competition between $\beta+$ decay and proton emission in the isomeric unbound nucleus sn 97. *Phys. Rev. C* **109**, L051302 (2024).
14. Adel, A., Mahmoud, K. H. & Taha, H. A. Exploring the competition between $\alpha$-decay and proton radioactivity: A comparative study of proximity potential formalisms. *Nucl. Phys. A* **1046**, 122872 (2024).

15. Zhang, D.-M. *et al.* Theoretical calculations of proton emission half-lives based on a deformed gamow-like model. *Chin. Phys. C* **48**, 044102 (2024).
16. Karny, M. *et al.* Shell structure beyond the proton drip line studied via proton emission from deformed 141ho. *Phys. Lett. B* **664**, 52–56 (2008).
17. Bugrov, V. & Kadmenskii, S. Proton decay of deformed nuclei. *Sov. J. Nucl. Phys. (English Transl.* **49** (1989).
18. Kadmensky, S. & Bugrov, V. Proton decay and shapes of neutron-deficient nuclei. *Phys. At. Nucl.* **59** (1996).
19. Cheng, J.-H. *et al.* Systematic study of proton radioactivity of spherical proton emitters with skyrme interactions. *The Eur. Phys. J. A* **56**, 273 (2020).
20. Qi, C., Delion, D. S., Liotta, R. J. & Wyss, R. Effects of formation properties in one-proton radioactivity. *Phys. Rev. C* **85**, 011303 (2012).
21. Deng, J.-G., Li, X.-H., Chen, J.-L., Cheng, J.-H. & Wu, X.-J. Systematic study of proton radioactivity of spherical proton emitters within various versions of proximity potential formalisms. *The Eur. Phys. J. A* **55**, 58 (2019).
22. Rykaczewski, K. New experimental results in proton radioactivity. *The Eur. Phys. J. A* **15**, 81–84 (2002).
23. Sarmiento, L. G. *et al.* Elucidating the nature of the proton radioactivity and branching ratio on the first proton emitter discovered 53mco. *Nat. Commun.* **14**, 5961 (2023).
24. Jackson, K., Cardinal, C., Evans, H., Jelley, N. & Cerny, J. 53com: A proton-unstable isomer. *Phys. Lett. B* **33**, 281–283 (1970).
25. Åberg, S., Semmes, P. B. & Nazarewicz, W. Spherical proton emitters. *Phys. Rev. C* **56**, 1762 (1997).
26. Basu, D., Chowdhury, P. R. & Samanta, C. Folding model analysis of proton radioactivity of spherical proton emitters. *Phys. Rev. C* **72**, 051601 (2005).
27. Balasubramaniam, M. & Arunachalam, N. Proton and $\alpha$-radioactivity of spherical proton emitters. *Phys. Rev. C* **71**, 014603 (2005).
28. Bhattacharya, M. & Gangopadhyay, G. Microscopic calculation of half lives of spherical proton emitters. *Phys. Lett. B* **651**, 263–267 (2007).
29. Dong, J., Zhang, H. & Royer, G. Proton radioactivity within a generalized liquid drop model. *Phys. Rev. C* **79**, 054330 (2009).
30. Hong-Fei, Z. *et al.* Theoretical analysis and new formulae for half-lives of proton emission. *Chin. Phys. Lett.* **26**, 072301 (2009).
31. Jian-Min, D., Hong-Fei, Z., Wei, Z. & Jun-Qing, L. Unified fission model for proton emission. *Chin. Phys. C* **34**, 182 (2010).
32. Qian, Y.-B., Ren, Z.-Z., Ni, D.-D. & Sheng, Z.-Q. Half-lives of proton emitters with a deformed density-dependent model. *Chin. Phys. Lett.* **27**, 112301 (2010).
33. Ferreira, L., Maglione, E. & Ring, P. Self-consistent description of proton radioactivity. *Phys. Lett. B* **701**, 508–511 (2011).
34. Ni, D. & Ren, Z. Coupled-channels study of fine structure in the $\alpha$ decay of platinum isotopes. *Phys. Rev. C* **84**, 037301 (2011).
35. Zhao, Q., Dong, J. M., Song, J. L. & Long, W. H. Proton radioactivity described by covariant density functional theory with the similarity renormalization group method. *Phys. Rev. C* **90**, 054326 (2014).
36. Qian, Y. & Ren, Z. Calculations on decay rates of various proton emissions. *The Eur. Phys. J. A* **52**, 1–7 (2016).
37. Zdeb, A., Warda, M., Petrache, C. & Pomorski, K. Proton emission half-lives within a gamow-like model. *The Eur. Phys. J. A* **52**, 1–6 (2016).
38. Santhosh, K. & Sukumaran, I. Description of proton radioactivity using the coulomb and proximity potential model for deformed nuclei. *Phys. Rev. C* **96**, 034619 (2017).
39. Gharaei, R., Shakib, M. J. & Santhosh, K. Description of temperature effects on proton radioactivity. *Nucl. Phys. A* **1037**, 122700 (2023).
40. Fiorin, G., Maglione, E. & Ferreira, L. Theoretical description of deformed proton emitters: nonadiabatic quasiparticle method. *Phys. Rev. C* **67**, 054302 (2003).
41. Delion, D., Liotta, R. & Wyss, R. Systematics of proton emission. *Phys. Rev. Lett.* **96**, 072501 (2006).
42. Medeiros, E., Rodrigues, M., Duarte, S. & Tavares, O. Systematics of half-lives for proton radioactivity. *The Eur. Phys. J. A* **34**, 417–427 (2007).
43. Qi, C., Delion, D. S., Liotta, R. J. & Wyss, R. Effects of formation properties in one-proton radioactivity. *Phys. Rev. C* **85**, 011303 (2012).
44. Sreeja, I. & Balasubramaniam, M. An empirical formula for the half-lives of ground state and isomeric state one proton emission. *The Eur. Phys. J. A* **54**, 106 (2018).
45. Zhang, Z.-X. & Dong, J.-M. A formula for half-life of proton radioactivity. *Chin. Phys. C* **42**, 014104 (2018).

46. Chen, J.-L. *et al.* New geiger-nuttall law for proton radioactivity. *The Eur. Phys. J. A* **55**, 214 (2019).
47. Soylu, A., Koyuncu, F., Gangopadhyay, G., Dehghani, V. & Alavi, S. Proton radioactivity half-lives with nuclear asymmetry factor. *Chin. Phys. C* **45**, 044108 (2021).
48. Denisov, V. Y. Empirical relations for $\alpha$-decay half-lives: The effect of deformation of daughter nuclei. *Phys. Rev. C* **110**, 014604 (2024).
49. Sarriguren, P., Algora, A. & Pereira, J. Gamow-teller response in deformed even and odd neutron-rich zr and mo isotopes. *Phys. Rev. C* **89**, 034311 (2014).
50. Santhosh, K. Theoretical studies on two-proton radioactivity. *Phys. Rev. C* **104**, 064613 (2021).
51. Budaca, R. & Budaca, A. Deformation dependence of the screened decay law for proton emission. *Nucl. Phys. A* **1017**, 122355 (2022).
52. Saxena, G. *et al.* Deformation dependence of 2p-radioactivity half-lives: probe with a new formula across the mass region with z< 82. *J. Phys. G: Nucl. Part. Phys.* **50**, 015102 (2022).
53. Dehghani, V. & Alavi, S. Empirical formulas for proton decay half-lives: Role of nuclear deformation and q-value. *Chin. Phys. C* **42**, 104101 (2018).
54. Ni, D. & Ren, Z. New formula of half-lives for proton emission from spherical and deformed nuclei. *continuum* **3**, 4 (2012).
55. Nilsson, S. & Damgaard, J. The present status in super-heavy element calculations. *Phys. Scripta* **6**, 81 (1972).
56. Eisenberg, J. M. & Greiner, W. Nuclear theory. microscopic theory of the nucleus. *ETDEWEB* (1976).
57. Aggarwal, M. Proton radioactivity at non-collective prolate shape in high spin state of 94ag. *Phys. Lett. B* **693**, 489–493 (2010).
58. Aggarwal, M. Coexisting shapes with rapid transitions in odd-z rare-earth proton emitters. *Phys. Rev. C* **89**, 024325 (2014).
59. Aggarwal, M., Saxena, G. & Parab, P. Correlation between the shape coexistence and stability in mo and ru isotopes. *Nucl. Phys. A* **1044**, 122843 (2024).
60. Lalazissis, G. *et al.* The effective force nl3 revisited. *Phys. Lett. B* **671**, 36–41 (2009).
61. Geng, L., Toki, H., Sugimoto, S. & Meng, J. Relativistic mean field theory for deformed nuclei with pairing correlations. *Prog. theoretical physics* **110**, 921–936 (2003).
62. Gambhir, Y., Ring, P. & Thimet, A. Relativistic mean field theory for finite nuclei. *Annals Phys.* **198**, 132–179 (1990).
63. Flocard, H., Quentin, P., Kerman, A. & Vautherin, D. Nuclear deformation energy curves with the constrained hartree-fock method. *Nucl. Phys. A* **203**, 433–472 (1973).
64. Ring, P. Relativistic mean field theory in finite nuclei. *Prog. Part. Nucl. Phys.* **37**, 193–263 (1996).
65. Singh, D., Saxena, G., Kaushik, M., Yadav, H. L. & Toki, H. Study of two-proton radioactivity within the relativistic mean-field plus bcs approach. *Int. J. Mod. Phys. E* **21**, 1250076 (2012).
66. Gamow, G. Zur quantentheorie des atomkernes. *Zeitschrift für Physik* **51**, 204–212 (1928).
67. Segre, E. Nuclei and particles: an introduction to nuclear and subnuclear physics. *(No Title)* (1964).
68. Deng, J.-G. & Zhang, H.-F. Correlation between $\alpha$-particle preformation factor and $\alpha$ decay energy. *Phys. Lett. B* **816**, 136247 (2021).
69. Kondev, F., Wang, M., Huang, W., Naimi, S. & Audi, G. The nubase2020 evaluation of nuclear physics properties. *Chin. Phys. C* **45**, 030001 (2021).
70. Saxena, G., Kumawat, M., Kaushik, M., Jain, S. & Aggarwal, M. Two-proton radioactivity with 2p halo in light mass nuclei a= 18–34. *Phys. Lett. B* **775**, 126–129 (2017).
71. Wang, N., Liu, M., Wu, X. & Meng, J. Surface diffuseness correction in global mass formula. *Phys. Lett. B* **734**, 215–219 (2014).
72. Xu, Y., Goriely, S., Jorissen, A., Chen, G. & Arnould, M. Databases and tools for nuclear astrophysics applications-brussels nuclear library (bruslib), nuclear astrophysics compilation of reactions ii (nacre ii) and nuclear network generator (netgen). *Astron. & Astrophys.* **549**, A106 (2013).
73. Möller, P., Sierk, A. J., Ichikawa, T. & Sagawa, H. Nuclear ground-state masses and deformations: Frdm (2012). *At. Data Nucl. Data Tables* **109**, 1–204 (2016).
74. Denisov, V. Y. & Khudenko, A. $\alpha$-decay half-lives: Empirical relations. *Phys. Rev. C* **79**, 054614 (2009).
75. Mukha, I. *et al.* Proton–proton correlations observed in two-proton radioactivity of 94ag. *Nature* **439**, 298–302 (2006).
76. Aggarwal, M. Shape coexistence in excited odd-z proton emitters eu 131- 136. *Phys. Rev. C* **90**, 064322 (2014).
77. Pakarinen, J. *et al.* Evidence for prolate structure in light pb isotopes from in-beam $\gamma$-ray spectroscopy of pb 185. *Phys. Rev. C* **80**, 031303 (2009).
78. Heenen, P.-H., Skalski, J., Staszczak, A. & Vretenar, D. Shapes and $\alpha$-and $\beta$-decays of superheavy nuclei. *Nucl. Phys. A* **944**, 415–441 (2015).

79. Yusofvand, Z. & Naderi, D. Calculation of the half-life of proton radioactivity using the empirical formula depending on angular momentum. *Int. J. Mod. Phys. E* **33**, 2450014 (2024).
80. Rodriguez, J. D., Perez, A. & Lozano, J. A. Sensitivity analysis of k-fold cross validation in prediction error estimation. *IEEE transactions on pattern analysis machine intelligence* **32**, 569–575 (2009).
81. Saxena, G., Sharma, P. & Saxena, P. A global study of $\alpha$-clusters decay in heavy and superheavy nuclei with half-life and preformation factor. *The Eur. Phys. J. A* **60**, 50 (2024).
82. Carnini, M. & Pastore, A. Trees and forests in nuclear physics. *J. Phys. G: Nucl. Part. Phys.* **47**, 082001 (2020).
83. Li, C.-Q., Tong, C.-N., Du, H.-J. & Pang, L.-G. Deep learning approach to nuclear masses and $\alpha$-decay half-lives. *Phys. Rev. C* **105**, 064306 (2022).
84. Zhang, H., Wang, Y., Dong, J., Li, J. & Scheid, W. Concise methods for proton radioactivity. *J. Phys. G: Nucl. Part. Phys.* **37**, 085107 (2010).
85. Guzmán, F. *et al.* Proton radioactivity from proton-rich nuclei. *Phys. Rev. C* **59**, R2339 (1999).
86. Guo, C., Zhang, G. & Le, X. Study of the universal function of nuclear proximity potential from density-dependent nucleon–nucleon interaction. *Nucl. Phys. A* **897**, 54–61 (2013).
87. Chen, J.-L., Li, X.-H., Cheng, J.-H., Deng, J.-G. & Wu, X.-J. Systematic study of proton radioactivity based on gamow-like model with a screened electrostatic barrier. *J. Phys. G: Nucl. Part. Phys.* **46**, 065107 (2019).
88. Wang, M., Huang, W. J., Kondev, F. G., Audi, G. & Naimi, S. The ame 2020 atomic mass evaluation (ii). tables, graphs and references. *Chin. Phys. C* **45**, 030003 (2021).
89. Möller, P., Mumpower, M. R., Kawano, T. & Myers, W. D. Nuclear properties for astrophysical and radioactive-ion-beam applications (ii). *At. Data Nucl. Data Tables* **125**, 1–192 (2019).
90. Lalazissis, G., Raman, S. & Ring, P. Ground-state properties of eveneven nuclei in the relativistic mean-field theory. *At. Data Nucl. Data Tables* **71**, 1–40, DOI: https://doi.org/10.1006/adnd.1998.0795 (1999).
91. Sobhani, H. & Khalafi, H. A comprehensive semi-empirical formula for the half-lives of beta-decaying nuclei. *Chin. J. Phys.* **85**, 475–507 (2023).
92. Wang, S. & Nazarewicz, W. Puzzling two-proton decay of kr 67. *Phys. Rev. Lett.* **120**, 212502 (2018).
93. Goigoux, T. *et al.* Two-proton radioactivity of kr 67. *Phys. Rev. Lett.* **117**, 162501 (2016).

## Acknowledgement

Authors are thankful to S.K. Jain, Manipal University Jaipur, Jaipur, India for the valuable communication. The support provided by SERB (DST), Govt. of India under CRG/2019/001851 grant is acknowledged. M. Aggarwal acknowledges the support provided by DST, Govt. of India. under SR/WOS-A/PM-30/2019 scheme.